\documentclass[aip,jcp,reprint]{revtex4-1}

\usepackage{amsmath}
\usepackage{graphicx}
\usepackage{dcolumn}
\usepackage{bm}
\usepackage{mathrsfs}
\usepackage{mhchem}
\usepackage{subfigure}
\usepackage{color}
\usepackage{ulem}

\draft

\begin{document}

\title{Single particle motion and collective dynamics in Janus motor systems}
\author{Mu-Jie Huang}
\email{mjhuang@chem.utoronto.ca}
\affiliation{ Chemical Physics Theory Group, Department of Chemistry, University of Toronto, Toronto, Ontario M5S 3H6, Canada}

\author{Jeremy Schofield}
\email{jmschofi@chem.utoronto.ca}
\affiliation{ Chemical Physics Theory Group, Department of Chemistry, University of Toronto, Toronto, Ontario M5S 3H6, Canada}

\author{Pierre Gaspard}
\email{gaspard@ulb.ac.be}
\affiliation{ Center for Nonlinear Phenomena and Complex Systems, Universit{\'e} Libre de Bruxelles (U.L.B.), Code Postal 231, Campus Plaine, B-1050 Brussels, Belgium}

\author{Raymond Kapral}
\email{rkapral@chem.utoronto.ca}
\affiliation{ Chemical Physics Theory Group, Department of Chemistry, University of Toronto, Toronto, Ontario M5S 3H6, Canada}

\date{\today}

\begin{abstract}
The single-particle and collective dynamics of systems comprising Janus motors, solvent and reactive solute species maintained in nonequilibrium states are investigated. Reversible catalytic reactions with the solute species take place on the catalytic faces of the motors, and  the nonequilibrium states are established by imposing either constant-concentration reservoirs that feed and remove reactive species, or through out-of-equilibrium fluid phase reactions. We consider general intermolecular interactions between the Janus motor hemispheres and the reactive species. For single motors, we show that the reaction rate depends nonlinearly on an applied external force when the system is displaced far from equilibrium. We also show that a finite-time fluctuation formula derived for fixed catalytic particles describes the nonequilibrium reactive fluctuations of moving Janus motors. Simulation of the collective dynamics of small ensembles of Janus motors with reversible kinetics under nonequilibrium conditions are carried out and the spatial and orientational correlations of dynamic cluster states are discussed. The conditions leading to the instability of the homogeneous motor distribution and the onset of nonequilibrium dynamical clustering are described.

\end{abstract}


\maketitle
\section{Introduction}

Active matter comes in a variety of shapes and forms ranging from fluids and gels to granular matter.~\cite{V12,A13,M16,R17} The agents that give rise to the activity are equally diverse and include living animals and organisms with macroscopic sizes to molecular machines with nanoscale dimensions.~\cite{J02} This article focuses on a specific class of active soft matter systems, namely, fluid systems where the active agents are synthetic motors that operate by diffusiophoretic mechanisms powered by chemical energy.~\cite{DD74,ALP82,A89,AP91,GLA05} In contrast to most molecular machines where chemical energy is used to drive conformational changes, the motors of interest in the present study have no moving parts and instead exploit asymmetric chemical activity on their surfaces to induce motion in solution.~\cite{K13} While these motors typically have sizes in the micrometer to nanometer range they may have different shapes, such as rods, spheres and sphere-dimers, as well as other more complex shapes.

Briefly, propulsion by self-diffusiophoresis arises as a result of chemical reactions that take place on a portion of the motor to produce local concentration gradients of reactant and product molecules. If these species interact with the motor through intermolecular potentials that differ from those of the solvent in which they reside, a net body force is produced that acts on the motor. Since momentum is conserved this force must be balanced by an equal and opposite force on the surrounding fluid. This gives rise to fluid flows that lead to propulsion.

Diffusiophoresis is not the only mechanism that can give rise to propulsion of synthetic motors without moving parts. Indeed, some of the first nanomotors were bimetallic rods propelled by electrophoresis,~\cite{PKOSACMLC04,FBAMO05} or by thermophoresis~\cite{A89,JYS10,YR11}, a propulsion mechanism that has many features in common with diffusiophoresis. In addition, in some cases an asymmetric chemical reaction can produce bubbles on the surface of a motor, which provide another type of propulsion mechanism.~\cite{W13}

Some basic principles and conditions must be satisfied for a full theoretical description of motor operation. At a microscopic level the fundamental time reversal symmetry of the dynamics must be preserved. As a consequence, detailed balance for reactive and nonreactive events must be satisfied. Since autonomous directed motion is not possible in a system at equilibrium, these motors can only operate if the system is displaced from equilibrium. Coupling of the system to reservoirs that control the concentrations of chemical species giving rise to a nonzero chemical affinity breaks detailed balance and leads to motor motion. Finally, micro/nanometer-sized motors are small so that thermal fluctuations cannot be neglected, necessitating microscopic or Langevin treatments of the dynamics that account for such fluctuations.

While studies of single-motor dynamics offer opportunities to probe details of the propulsion mechanisms and study scenarios for motor control and cargo transport, the nature of the collective dynamics of many-motor systems presents new phenomena such as the emergence of active self-assembly into various kinds of dynamical cluster states. The complex physics of motors interacting both directly and indirectly through hydrodynamic flows and chemotaxis also poses formidable new theoretical challenges. Key points that need to be addressed include unraveling the relative importance of various concentration and fluid flow coupling effects among motors that are responsible for observed nonequilibrium system states, and the construction of theories that can describe correlations in active nonequilibrium systems.

In this article we consider the dynamics of individual Janus motors as well as their collective dynamics. While there have been other experimental and theoretical investigations of Janus motor systems,~\cite{KYCS10,TCPYB12,SS12,dBK13,MHS13,GPFHW13,GPDW14,YWR14,EGDHILG14,HSK16,OPD17,HSK17,CEIG18,HSGK18} in this study we emphasize some features of the descriptions of Janus motors that are often not fully considered. In particular, both the theoretical and simulation methods employed in this study satisfy detailed balance at equilibrium and nonequilibrium conditions are established either by constant-concentration reservoirs or by fluid phase reactions that, themselves, take place out of equilibrium. This allows us to specify the affinities that characterize the nonequilibrium system states for both the single-motor and many-motor systems.

The outline of the paper is as follows: Section~\ref{sec:self-diff} presents an overview of the Langevin equations for the linear and angular velocities of a single motor and the reaction rate that follow from a fluctuating thermodynamics description of the system. The results in this section form the basis for the analysis of the simulation results in Sec.~\ref{sec:single-Janus}. The model for the Janus motor and its environment is the same as that described that in an earlier study.~\cite{HSGK18} New simulation results on the dependence of the reaction rate on an applied external force and the applicability of the finite-time fluctuation formula for diffusion-influenced reactions to self-propelled Janus motors are given, as well as simulation results for systems where the reactive species have different interactions on the catalytic and noncatalytic hemispheres of the Janus colloid. Collective motion is the topic of Sec.~\ref{sec:collective} where simulation results are given and discussed. The conclusion and perspectives of the study are given in Sec.~\ref{sec:conc} and the Appendices provide additional technical details.

\section{Self-diffusiophoretic Janus motors}\label{sec:self-diff}
\subsection{Langevin description}
We begin with a short overview of the Langevin description of the dynamics and reaction rate for a Janus colloidal motor in solution since this formulation will be used to interpret the simulation results on single-motor motion and extended in applications to collective motor dynamics.

We consider the dynamics of a spherical Janus colloidal motor with catalytic ($C$) and noncatalytic ($N$) hemispherical caps whose orientation is specified by a unit vector ${\bf u}$ directed along the polar axis from the $N$ to $C$ hemispheres. The motor has a radius $R$ whose value lies in either the micron or sub-micron range so that fluctuations play a significant role in the dynamics of the motor. The Janus colloid is immersed in a fluid containing chemically inert solvent $S$ and solute $k=A,B$ species. Chemical reactions with forward and reverse rate constants per unit area $\kappa_\pm$, $C+A \underset{\kappa_-}{\stackrel{\kappa_+}{\rightleftharpoons}} C+B$, take place on the catalytic cap and power motor motion by a self-diffusiophoretic mechanism. Langevin equations for the linear ${\bf V}$ and angular $\pmb{\Omega}$ velocities of a Janus motor have been derived using a fluctuating thermodynamics formulation that accounts for the fluctuating concentration and velocity fields in the bulk of the solution and fluctuating boundary conditions on the surface of the colloidal particle.\cite{GK17,GK18a,GK18b} The Langevin equations are
\begin{eqnarray}
M\frac{d{\bf V}}{dt} &=& -\gamma_{\rm t}\, {\bf V}  + {\bf F}_{\rm sd} + {\bf F}_{\rm ext} + {\bf F}_{\rm fl}(t) \, ,
\label{Langevin-eq-trans}\\
{\boldsymbol{\mathsf I}}\cdot\frac{d\pmb{\Omega}}{dt} &=& - \gamma_{\rm r}\,\pmb{\Omega} + {\bf T}_{\rm sd} + {\bf T}_{\rm ext} + {\bf T}_{\rm fl}(t) \, ,
\label{Langevin-eq-rot}
\end{eqnarray}
where $M$ and ${\boldsymbol{\mathsf I}}$ denote the mass and inertia tensor of the motor, the translational friction coefficient is $\gamma_{\rm t}= 6\pi\eta R (1+2b/R)/(1+3b/R)$ and the rotational friction coefficient is $\gamma_{\rm r}=8\pi\eta  R^3/(1+3b/R)$. Both of these coefficients are written for partial slip boundary conditions for the fluid velocity field on the surface of the colloid. The slip length $b=\eta/\lambda$ is expressed as the ratio of the fluid viscosity $\eta$ and the coefficient of sliding friction $\lambda$. We suppose that the slip length is uniform on the whole surface, assuming that the interaction of the motor with the solvent molecules is uniform and dominates over the motor interactions with the solute molecules. The Gaussian random force and torque, ${\bf F}_{\rm fl}$ and ${\bf T}_{\rm fl}$, both have zero mean and satisfy fluctuation-dissipation relations $\langle {\bf F}_{\rm fl}(t)\, {\bf F}_{\rm fl}(t')\rangle = 2k_{\rm B}T \, \gamma_t \, \delta(t-t') \, {\boldsymbol{\mathsf 1}}$ and $\langle {\bf T}_{\rm fl}(t)\, {\bf T}_{\rm fl}(t')\rangle = 2k_{\rm B}T \gamma_{\rm r} \, \delta(t-t') \, {\boldsymbol{\mathsf 1}}$, respectively, where {\boldsymbol{\mathsf 1}} denotes the unit tensor. The position $\bf r$ and orientation unit vector $\bf u$ of the Janus motor are obtained by integrating the evolution equations ${\bf V}=d{\bf r}/dt$ and $d{\bf u}/dt=\pmb{\Omega}\times{\bf u}$ with respect to time.

The terms that distinguish these Langevin equations from standard Langevin equations for inactive Brownian particles are the diffusiophoretic force and torque contributions,
\begin{eqnarray}
{\bf F}_{\rm sd}(t) &=& \frac{6\pi\eta R}{1+3b/R} \sum_{h=C}^N\sum_{k=A}^B  b^h_k \, \overline{H_h(\theta)\pmb{\nabla}_\theta c_k({\bf r},t)}^{\rm s} ,\label{Fd-surf-av} \\
{\bf T}_{{\rm sd}}(t)&=& \frac{12\pi\eta  R}{1+3b/R} \, \sum_{h=C}^N\sum_{k=A}^B  b^h_k  \, \overline{H_h(\theta){\bf r}\times\pmb{\nabla}c_k({\bf r},t)}^{\rm s} , \label{rot-diff-t}
\end{eqnarray}
where $H_h(\theta)$ is a Heaviside function that is unity on the $h=C,N$ hemisphere and zero otherwise, and $\pmb{\nabla}_\theta$ is the gradient in the direction tangential to the colloidal surface. These functions
depend on the surface averages, $\overline{(\cdot)}^s=(4\pi R^2)^{-1}\int_S(\cdot)dS$, of the products of the tangential gradients of the local concentrations of species $k$, and the diffusiophoretic coefficients, $b^h_k = \frac{k_{\rm B}T}{\eta} \big( K_k^{h(1)} + b\, K_k^{h(0)}\big)$. The $K_k^{h(n)}$ are defined in terms of integrals over the interface of factors that depend in the interactions potentials $u^h_k$ of the solute species $k$ with the cap $h$ of the colloid: $K_k^{h(n)} \equiv \int dz \, z^n \, \big[ {\rm e}^{-\beta u^h_k(z)}-1 \big]$, where the integration is performed in the $z$-direction normal to the surface.~\cite{A89,AP91,AB06} Note that the diffusiophoretic force and torque are independent of the fluid viscosity in view of the definition of $b_k$ and are finite in the limits of perfect stick ($b=0$) and perfect slip ($b \to \infty$). The self-diffusiophoretic torque vanishes by symmetry for a single Janus colloid, although an active torque with the form in Eq.~(\ref{rot-diff-t}) can arise in the presence of external concentration gradients, and this effect will enter when collections of active Janus particles are considered in Sec.~\ref{sec:collective}.

For micron and submicron-sized Janus motors in solution the P\'{e}clet number is typically small and the reaction-diffusion equations for the concentration fields effectively decouple from those describing the fluid flow velocity.  In this low P\'{e}clet number regime, the concentration fields that enter the Langevin equations are the solutions of the steady-state diffusion equations, $D_k \nabla^2 c_k =0$, which must be solved subject to the radiation boundary conditions, $D_k \partial_r c_k =-\nu_k w H_C$ on the surface $r=R$ of the Janus colloid. Here the reaction rate is $w= \kappa_+c_A - \kappa_-c_B$, $\nu_k$ is the stoichiometric coefficient of species $k$ (negative for reactants and positive for products) and $\kappa_{\pm}$ are the surface reaction rate constants. When the system is maintained in a nonequilibrium state by fixing the concentrations at $c_k=\bar{c}_k$ in reservoirs far from the motor, the dimensionless chemical affinity that characterizes the nonequilibrium state is defined as $A_{\rm rxn} \equiv \ln (\kappa_+\bar{c}_A/\kappa_-\bar{c}_B)$.

Under low Reynolds number conditions, the inertial terms on the left sides of Eqs.~(\ref{Langevin-eq-trans}) and (\ref{Langevin-eq-rot}) can be neglected and the resulting overdamped equations are,
\begin{eqnarray}\label{eq:overdamped}
\frac{d{\bf r}}{dt} &=& \frac{1}{\gamma_{\rm t}} \, {\bf F}_{\rm ext} + {\bf V}_{\rm sd}  + {\bf V}_{\rm fl}(t) \\
\frac{d{\bf u}}{dt} &=& -\frac{1}{\gamma_{\rm r}} \, {\bf u}\times \left[{\bf T}_{\rm ext} + {\bf T}_{\rm fl}(t) \right],
\end{eqnarray}
where the self-diffusiophoretic velocity is ${\bf V}_{\rm sd}=V_{\rm sd} {\bf u}={\bf F}_{\rm sd}/\gamma_{\rm t}$ with an analogous expression for the fluctuating velocity, ${\bf V}_{\rm fl}(t)$, derived from the fluctuating force.

Since chemical reactions on the motor surface drive propulsion, we must also consider the Langevin equation for the reaction rate. The mean reaction rate is the integration of $w$ over the catalytic surface of the Janus particle, $W_{\rm rxn}=\int_{S_C}dS \; w =\Gamma (\kappa_+\bar{c}_A -  \kappa_-\bar{c}_B)$, where the last equality expresses the mean rate in terms of the fixed concentrations in the reservoirs. The prefactor $\Gamma$ depends on the solution for the concentration field. The net rate of production of product molecules in a time interval $[0,t]$, $dn/dt=dN_B/dt=-dN_A/dt$, where $N_k$ is the number of molecules of species $k$, satisfies the stochastic equation,
\begin{equation}
\frac{dn}{dt} = W_{\rm sd} + W_{\rm rxn} + W_{\rm fl}(t) \, ,
\label{eq-n}
\end{equation}
where the random reaction rate with zero mean accounts for fluctuations in the surface reaction rate and satisfies the fluctuation-dissipation relation, $\langle W_{\rm fl}(t)\, W_{\rm fl}(t')\rangle = 2D_{\rm rxn} \, \delta(t-t')$ with a reaction diffusivity given by $D_{\rm rxn}=(\Gamma/2)(\kappa_+\bar{c}_A +  \kappa_-\bar{c}_B)$. The diffusiophoretic contribution $W_{\rm sd}$ must be present for the formulation to be consistent with microscopic reversibility and takes the form,~\cite{GK17,GK18a} $W_{\rm sd}=\beta\chi D_{\rm rxn} {\bf u}\cdot{\bf F}_{\rm ext}$, where $\beta=1/(k_{\rm B}T)$ and the diffusiophoretic parameter $\chi$ is defined by $\chi=V_{\rm sd}/W_{\rm rxn}$.

\subsection{Spherical Janus motors}
For simplicity, we henceforth consider systems with equal species diffusion coefficients, $D_A=D_B=D$. Under this condition, the total concentration field, $c=c_A+c_B=c_0$, is constant as can be seen from the solution of the diffusion equation $\nabla^2 c =0$, with boundary condition $\partial_r c =0$ at $R$ and $c=\bar{c}_A+\bar{c}_B=c_0$ at infinity. The solution of the diffusion equation for $c_B$ can be written as an expansion in Legendre polynomials
\begin{equation}
c_B = \bar{c}_B + (\kappa_+ \bar{c}_A - \kappa_- \bar{c}_B)\frac{R}{D} \sum_{\ell=0}^{\infty} a_{\ell} f_{\ell}(r) P_{\ell}(\cos \theta),
\label{eq:cb_solution}
\end{equation}
where the solutions $f_{\ell}(r)$  of the radial equation and the expansion coefficients $a_{\ell}$ are given in Appendix~\ref{app:conc}. The parameter $\Gamma$ introduced above that enters the mean reaction rate $W_{\rm rxn}$ and the reaction diffusivity $D_{\rm rxn}$ is given explicitly by $\Gamma=4\pi R^2 a_0$.

From Eq.~(\ref{Fd-surf-av}) for the diffusiophoretic force, the self-diffusiophoretic velocity of the Janus motor for general interactions of the solute species with the Janus hemispheres may be written explicitly as
\begin{equation} \label{eq:Vu-cont}
{\bf V}_{\rm sd}= - \frac{k_{\rm B}T}{\eta D} \frac{\kappa_+ \bar{c}_A-\kappa_- \bar{c}_B}{2(1+2b/R)} \sum_{h=C}^N \Lambda_h \alpha_h \;{\bf u}=V_{\rm sd} \; {\bf u},
\end{equation}
where $\frac{k_{\rm B}T}{\eta}\Lambda_h= \frac{k_{\rm B}T}{\eta}(\Lambda_h^{(1)} +b \Lambda_h^{(0)})=b_B^h-b_A^h$ and
\begin{equation} \label{eq:alpha_h}
\alpha_h= \sum_{\ell=0}^\infty a_{\ell} \int_0^\pi d\theta \; H_h(\theta) \sin^2 \theta \;\partial_\theta P_{\ell}(\cos \theta).
\end{equation}
If $\Lambda_C=\Lambda_N=\Lambda$ so that the interactions of the $A$ and $B$ species with the motor do not depend on the identities of the hemispherical caps, the formula for the velocity reduces to~\cite{HSGK18} ${\bf V}_{\rm sd}=  \frac{2}{3}\frac{k_{\rm B}T}{\eta} \frac{(\kappa_+ \bar{c}_A-\kappa_- \bar{c}_B)}{D(1+2b/R)} \Lambda a_1 {\bf u}$, since \\ $(\alpha_C+\alpha_N)/2=-2a_1/3$.

The self-diffusiophoretic motion of the Janus particle is accompanied by fluid flow fields that form an integral part of the propulsion mechanism.~\cite{A86,A89,RHSK16,CEIG18} Accounting for partial slip of the fluid velocity field on the surface of the motor, as well as general interaction potentials, these fields are given by (see Appendix~\ref{app:v_field}) $\mathbf{v} = v_r \hat{{\bf r}} + v_{\theta} \hat{\bm{\theta}}$ with $\hat{{\bf r}}$ and $\hat{\bm{\theta}}$ unit vectors normal and transverse to the colloid surface, where
\begin{eqnarray}
v_r(r, \theta) &=& V_{\rm sd}( R/r)^3 P_1(\mu)  \\
&&+ \sum_{\ell= 2}^{\infty} (\ell + 1)\big[ ( R/r)^{\ell}  - ( R/r)^{\ell+2} \big] \chi_{\ell} P_{\ell} (\mu) , \label{eq:vr}\nonumber\\
v_{\theta}(r, \theta) &=& -\frac{V_{\rm sd}}{2}( R/r)^3 P_1^1(\mu) \\
&&+ \sum_{\ell= 2}^{\infty} \Big[\Big( \frac{2-\ell}{\ell}\Big) ( R/r)^{\ell} + (R/r)^{\ell+2} \Big] \chi_{\ell} P_{\ell}^1 (\mu) \label{eq:vt}.\nonumber
\end{eqnarray}
In Eqs. (\ref{eq:vr}) and (\ref{eq:vt}),  $\mu=\cos\theta$, $P_{\ell}^1 (\mu)$ is an associated Legendre polynomial, and $\chi_{\ell} = - \frac{1+2b/R}{1+(2\ell+1)b/R}\frac{2\ell+1}{2(\ell+1)} \sum_{m=1}^{\infty}B_{\ell m} a_{m}$ for $\ell \geq 2$ with
\begin{equation}
B_{\ell m} =\frac{k_{\rm B}T}{\eta D} \frac{\kappa_+ \bar{c}_A - \kappa_-\bar{c}_B}{2(1+2b/R)} \sum_{h=C}^N \Lambda_h  \int_{-1}^1  d\mu H_h P_{\ell}^1(\mu)  P_{m}^1(\mu) .
\label{eq:B_lm}
\end{equation}
Moreover, the self-diffusiophoretic torque~(\ref{rot-diff-t}) is vanishing for this Janus motor in a constant concentration background.

\section{Dynamics of single Janus motors}\label{sec:single-Janus}

We now turn to coarse-grained microscopic simulations of the dynamics of single Janus motors under nonequilibrium conditions established by coupling the system to reservoirs with fixed concentrations of chemical species as described above, as well as through bulk fluid phase reactions operating under nonequilibrium conditions. The equations of motion combine molecular dynamics of the Janus colloid interacting with the fluid species through intermolecular potentials and multiparticle collision dynamics~\cite{MK99,*Mk00} for the fluid particles. In contrast to the fluctuating thermodynamics formulation in the previous section where transport properties and boundary conditions are specified, in the microscopic simulations all transport and other properties follow from the dynamics and are determined by the intermolecular potentials and multiparticle collision parameters. Thus, comparisons between these two treatments of the dynamics provide contrasting perspectives on the Janus motor dynamics.

Specifically, we use a model that was constructed and studied previously~\cite{HSGK18} for a Janus motor made from small catalytic and noncatalytic spheres (beads) immersed in a fluid containing inert solvent and reactive $A$ and $B$ species. For this model reversible reactions $A\rightleftharpoons B$, which take place on the catalytic surface, were shown to satisfy microscopic reversibility. The reaction rates and Janus dynamics were investigated under equilibrium and nonequilibrium conditions. We use the same model here and refer the reader to Ref.~[\onlinecite{HSGK18}] for further details about the model and its implementation. The parameters used in this study are collected in Appendix~\ref{app:sim}. In the following mass is expressed in units of $m$, length in units of $\sigma$, energies in units of $k_{\rm B}T$ and time in units of $t_0 = \sqrt{m \sigma^2/k_{\rm B}T}$.

\subsection{Motors without fluid phase reactions}

We first present results on two aspects of the dynamics of Janus motors that have not been considered in the earlier investigations using this model. The simulation results described here are for systems with interaction potentials and reaction rate constants that satisfy $u^C_k=u^N_k$ and $\kappa_+ = \kappa_-$, respectively. Accordingly, it follows that $\Lambda_C=\Lambda_N=\Lambda$. The nonequilibrium state is established by fixing the concentrations in a reservoir at distance $r=L$ far from the Janus colloid and is controlled by the value of the chemical affinity $A_{\rm rxn}$.

The two aspects we investigate relate to the dependence of the reaction rate on the external force and the validity of the finite-time fluctuation formula for Janus motors. In both instances we investigate phenomena outside the expected domain of validity of the theories that describe these effects.

\subsubsection{Reaction rate and external force}

\begin{figure}[htbp]
\centering
\resizebox{1.0\columnwidth}{!}{%
\includegraphics{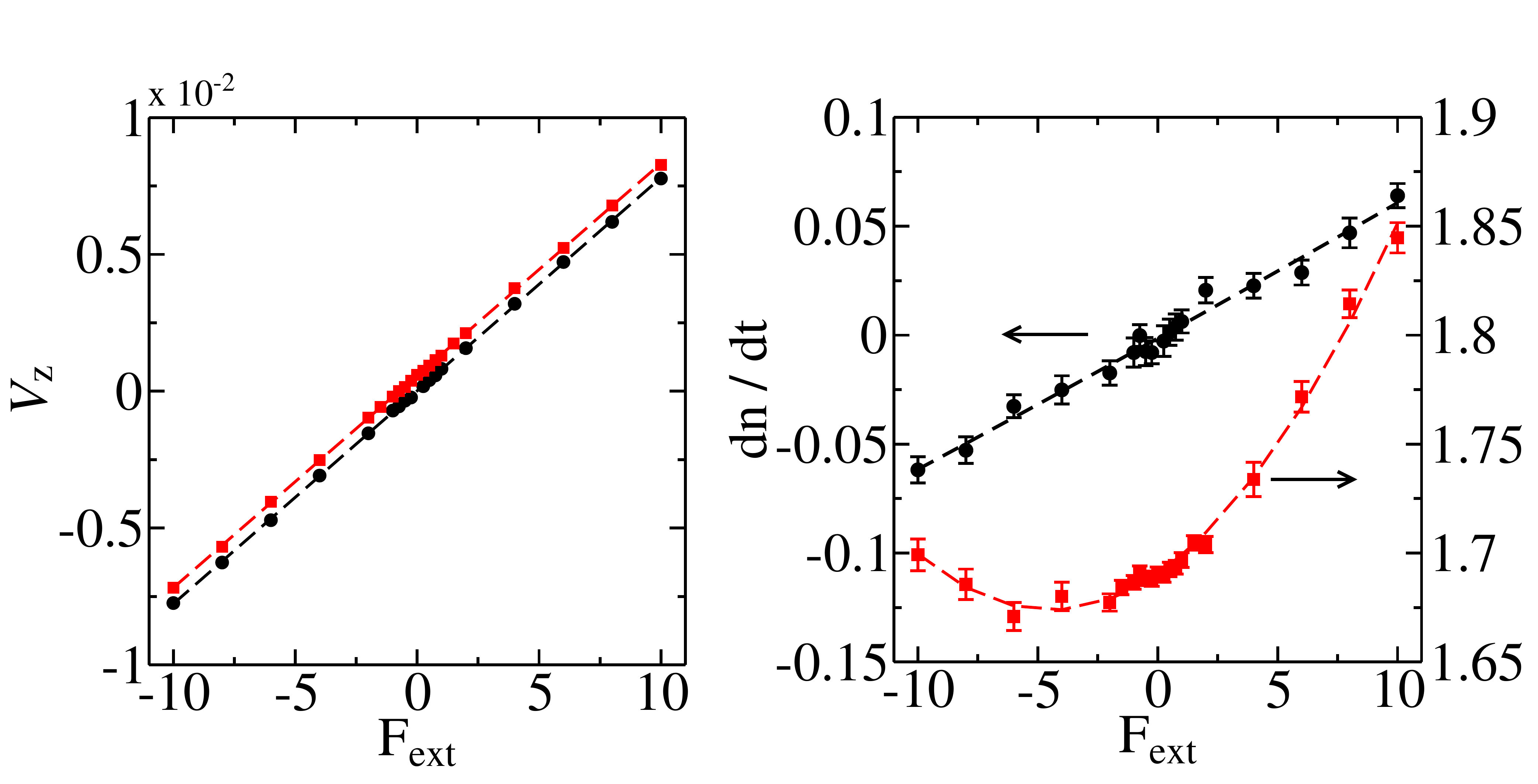}}
\caption{\label{fig:n_t_compare_nonlinear}Average motor velocity along the polar axis $\bf u$, $V_{\rm z}$  (left panel), and reaction rate, $dn/dt$ (right panel) versus external force with strength $\rm{F_{ext}}$, where the results are from systems with (a) $\bar{c}_A =\bar{c}_B = 10$ (black circles) and (b) $\bar{c}_A = 10$ and $\bar{c}_B = 9$ (red squares). The fits to the data presented in the text are plotted as dashed lines. }
\end{figure}

An interesting consequence of microscopic reversibility is the prediction that the reaction rate of diffusiophoretic motors depends on the applied external force.~\cite{GK17,GK18a} An analogous dependence of the reaction rate on the applied force is well known for biological molecular motors~\cite{JAP97,GG10} but has not been studied previously for synthetic diffusiophoretic motors. The fluctuating thermodynamics treatment outlined in Sec.~\ref{sec:self-diff} is limited to the linear regime and particle-based simulations have confirmed its validity within this linear regime.~\cite{HSGK18} Restriction to the linear regime can be controlled by adjusting the chemical and mechanical affinities,  $A_{\rm rxn}$ and ${\bf A}_{\rm mech}= \beta {\bf F}_{\rm ext}$, respectively. However, it is interesting to determine the domain of validity of this theory and examine how the reaction rate varies for higher values of the chemical and mechanical affinities.

To do this, as in our earlier study~\cite{HSGK18}, we suppose the motor has a magnetic moment and subject it to an external force $\mathbf{F}_{\rm ext} = F_{\rm ext} \hat{\mathbf{z}}$, as well as an external torque derived from a magnetic field in the same $z$ direction as the applied force, $\mathbf{B} = B\:\hat{\mathbf{z}}$, which controls the mean orientation of the motor at $\langle u_z \rangle=0.998$. We consider two systems with the same chemical affinities used to construct Fig.~6 of Ref.~[\onlinecite{HSGK18}], $A_{\rm rxn} = 0$ ($\bar{c}_A = \bar{c}_B=10$) and $A_{\rm rxn} \approx 0.1$ ($\bar{c}_A = 10$ and $\bar{c}_B=9$), but over a much larger range of external force values, $[-10,10]$. The results are plotted in Fig~\ref{fig:n_t_compare_nonlinear}.

The case of $A_{\rm rxn} = 0$ where the system is in chemical equilibrium in the absence of an external force is of interest since the diffusiophoretic constant is nonzero and finite even in the absence of a chemical driving force. For $A _{\rm rxn} = 0$ we see in Fig.~\ref{fig:n_t_compare_nonlinear} that $\langle dn/dt\rangle$ varies linearly with the external force and that negative values of $F_{\rm ext}$ correspond to $\langle dn/dt\rangle <0$ indicating that the motor consumes product and produces fuel.~\cite{HSGK18} A quadratic fit to the data yields $\langle dn /dt \rangle= -0.0015 + 0.006 F_{\rm ext} + 1\times 10^{-5} F_{\rm ext}^2 $, showing that the coefficient of the quadratic term is approximately two orders of magnitude smaller than the linear coefficient. For even larger values of $F_{\rm ext}$ the deviations from linearity are more evident. In contrast, for $A_{\rm rxn}=0.1$ strong deviations from linearity are observed with $\langle dn /dt \rangle= 1.69 + 0.0076 F_{\rm ext} + 0.001 F_{\rm ext}^2 $. In both cases, the self-diffusiophoretic motor velocity varies linearly with external force for both values of the chemical affinity over the entire external force range with slope $\gamma_{\rm t}^{-1} = 7.7\times 10^{-4}$.

This result shows that nonlinear effects manifest themselves in the reaction rate if the external force is large enough.  Such effects can be investigated by considering higher-order corrections in the P\'eclet number.  They are also ruled by microreversibility, establishing relationships between the nonlinear response coefficients and the statistical cumulants of the fluctuations in motion and reaction.\cite{AG04,AG07}

\subsubsection{Time-dependent fluctuation formula}

Recently it was shown that finite-time fluctuation formulas can be derived for diffusion-influenced surface reactions on fixed catalytic particles.~\cite{GK18c} These formulas, which hold for all times $t$,  take the form $P(n,t)/P(-n,t)=\exp{(\mathcal{A}_t n)}$, where $P(n,t)$ is the probability that a net number $n$ of product molecules is produced in the time interval $[0,t]$, and $\mathcal{A}_t$ is a time dependent affinity. The time dependence of $\mathcal{A}_t$ has its origin in the diffusive transport to the catalytic surface in the time interval $t$. In the limit of $t \to \infty$ the time-dependent affinity converges to $A_{\rm rxn}$ introduced earlier.

Specifically, for a fixed colloidal particle, the time dependence of the affinity is given by $\mathcal{A}_t = \ln (W_t^{(+)}/W_t^{(-)})$ through time dependence of the reaction rates,
\begin{eqnarray}
W_t^{(+)} &=& 2\pi R^2 \kappa(1-{\rm Da}\:\gamma_J)\bar{c}_A + \frac{c_0}{4}\, \frac{\Upsilon(t)}{t} \, ,  \\
W_t^{(-)} &=& 2\pi R^2 \kappa(1-{\rm Da}\:\gamma_J)\bar{c}_B + \frac{c_0}{4}\, \frac{\Upsilon(t)}{t}\, ,
\end{eqnarray}
written here for $D \equiv D_{A,B}$ and $\kappa \equiv \kappa_{\pm}$, again with $c_0=\bar{c}_A+\bar{c}_B$. The Damk{\"{o}}hler number ${\rm Da} = 2\kappa R/D$ characterizes the diffusion-influenced surface reaction and $\gamma_J = (1-2a_0)/{\rm Da}$ with $a_0$ the coefficient defined in Eq.~(\ref{eq:cb_solution}). The time-dependent function $\Upsilon(t)$ is expressed as
\begin{equation}
\Upsilon(t) = 4\pi \int_R^L dr \, \sum_{\ell=0}^{\infty} \frac{1}{2\ell +1} v_{\ell}(r,0) [ v_{\ell}(r,0) - v_{\ell}(r,t)],
\end{equation}
where $v_{\ell}(r,t)$ is the solution of $\partial_t v_{\ell}(r,t) = D[\partial_r^2 - \ell(\ell+1)/r^2] v_{\ell}(r,t)$ subject to the boundary conditions $(v_{\ell})_L = 0$ at $r=L$, and on motor surface at $r=R$,
\begin{eqnarray}
R(\partial_r v_{\ell})_R &=& (v_{\ell})_R \\
&& +\, {\rm Da}\, \frac{2\ell+1}{2} \sum_{m=0}^{\infty} (v_m)_R \int_0^1 d\mu P_{\ell}(\mu) P_m(\mu),\nonumber
\end{eqnarray}
with the initial condition, expressed in terms of the expansion coefficients $a_{\ell}$ of the concentration field,
\begin{equation}
v_{\ell}(r,0) = {\rm Da}\:R\:a_{\ell} \bigg[ \bigg(\frac{R}{r}\bigg)^{\ell} - \bigg(\frac{R}{L}\bigg)\bigg(\frac{r}{L}\bigg)^{\ell+1}  \bigg].
\end{equation}

As noted above, these results were derived for a fixed catalytic particle and have been confirmed by Langevin and coarse-grained microscopic simulations of the dynamics for fully catalytic and Janus spherical colloidal particles without propulsion.~\cite{GGHK18} Here, we consider self-propelled Janus colloids in order to determine the extent to which these finite-time fluctuation formulas might apply to these motors.

\begin{figure}[htbp]
\centering
\resizebox{0.8\columnwidth}{!}{%
\includegraphics{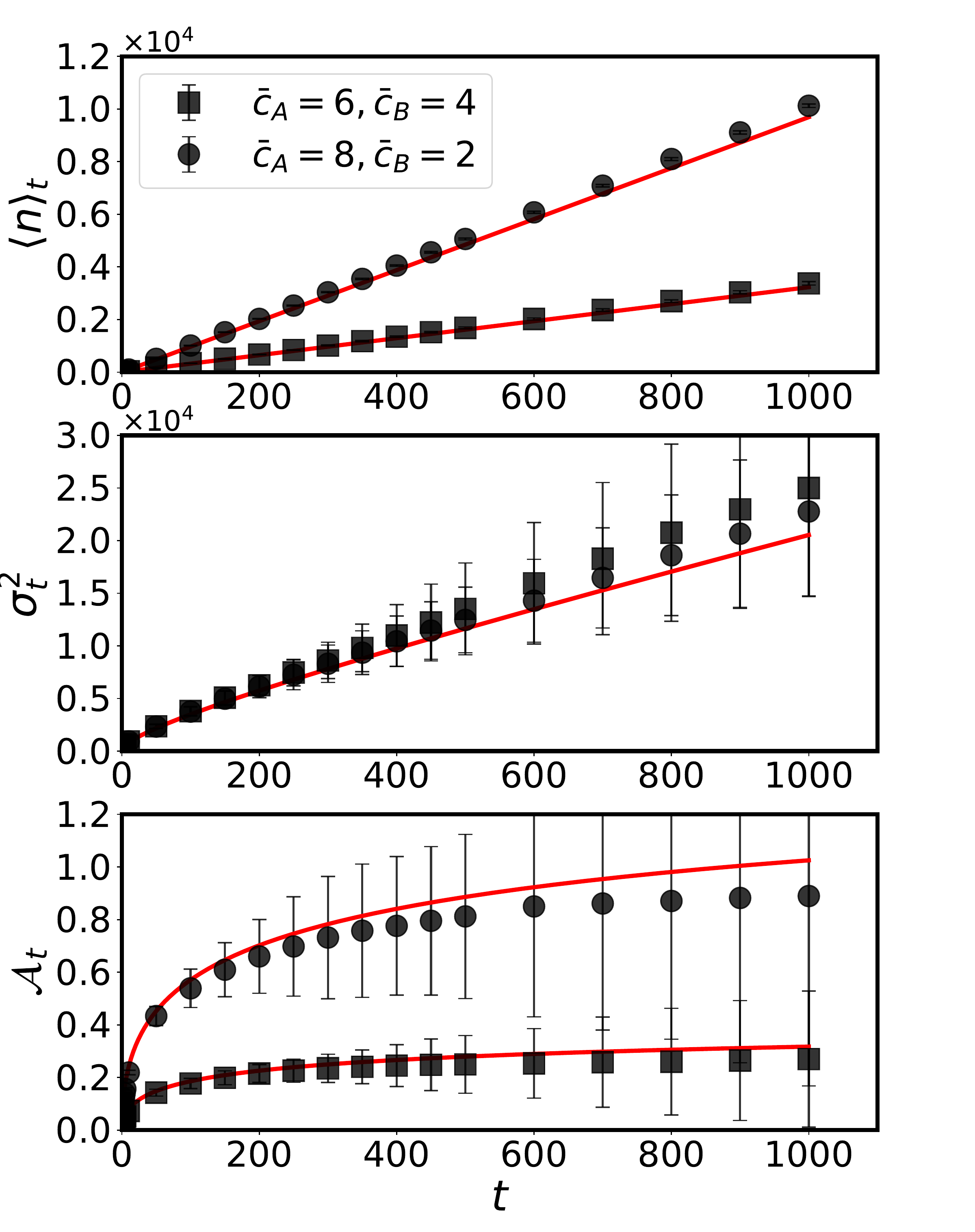}}
\caption{\label{fig:Janus_motor_forward}Plots of time-dependent mean number of reactive events, $\langle n \rangle_t$, the corresponding variance $\sigma_t^2$, and the affinity, $\mathcal{A}_t$ for Janus motors with propulsion driven by nonequilibrium boundary concentrations $\bar{c}_A = 6$ and $\bar{c}_A = 4$ (dots) ($V_{\rm sd}=0.0003$), and $\bar{c}_A = 8$ and $\bar{c}_A = 2$ (squares) ($V_{\rm sd}=0.0009$) with $\Lambda=\Lambda^{(1)}+b \Lambda^{(0)}= 0.122+b \, 0.0255=0.390$ with $b=10.5$. The red curves are theoretical predictions.}
\end{figure}

To estimate the time-dependent affinity in the Janus motor simulations, the probability distribution $P(n,t)$ is determined and fitted to the Gaussian probability distribution
\begin{equation}
P(n,t) = \frac{1}{\sqrt{2\pi \sigma_t^2}} \exp \bigg[\frac{-(n-\langle n \rangle_t)^2}{2\sigma_t^2}\bigg],
\end{equation}
where $\langle n \rangle_t = \mathcal{J}_t\:t$ and $\sigma_t^2 = 2{\cal D}_t\:t$ are the average number of reactive events in the time interval $t$ and the variance, respectively, written in terms of $\mathcal{J}_t = W_t^{(+)} - W_t^{(-)}$, and the reaction diffusivity, $\mathcal{D}_t = \frac{1}{2} (W_t^{(+)} + W_t^{(-)}) $. The affinity can be estimated as $\mathcal{A}_t \approx \mathcal{J}_t/ \mathcal{D}_t = 2\langle n \rangle_t / \sigma_t^2$. In Figure~\ref{fig:Janus_motor_forward} the simulation and the theoretical results are compared and good agreement is seen. Compared with the simulations without self-diffusiophoretic propulsion (i.e., with $\Lambda=0$),\cite{GGHK18}, the rate here is about $1.4\%$ larger with propulsion and the variance about $3.3\%$ higher, so that the affinity is roughly $2\%$ smaller. These results suggest that if the motor velocity is not too large a finite-time fluctuation formula can be used to analyze the statistics of the motor reactive events in nonequilibrium steady states.

\subsection{Motors with fluid phase reactions}\label{subsec:bulk}

Biological molecular machines operate in nonequilibrium environments controlled by chemical networks that produce motor fuel and consume motor products, often ATP and ADP, respectively. Similarly, nonequilibrium conditions for motor motion can also be established by fluid phase reactions whose mechanisms involve motor reactant and product species.~\cite{HSGK18} These reactions or reaction networks are themselves driven into nonequilibrium states by fluxes of other species whose values may be incorporated into effective rate constants. In this way detailed balance is broken so that directed motor motion is possible.  In the studies of collective dynamics presented in Sec.~\ref{sec:collective}, it is convenient to establish nonequilibrium conditions in this way without considering details of how fuel is supplied to the system at the boundaries. Consequently, we now present information on single Janus motor motion under these nonequilibrium conditions that will be needed later. We also consider the full generality of the intermolecular interactions of the solute species with the motor, allowing for different interaction potentials of these species with the catalytic and noncatalytic caps of the motor.

\begin{figure}[htbp]
\centering
\resizebox{0.9\columnwidth}{!}{%
\includegraphics{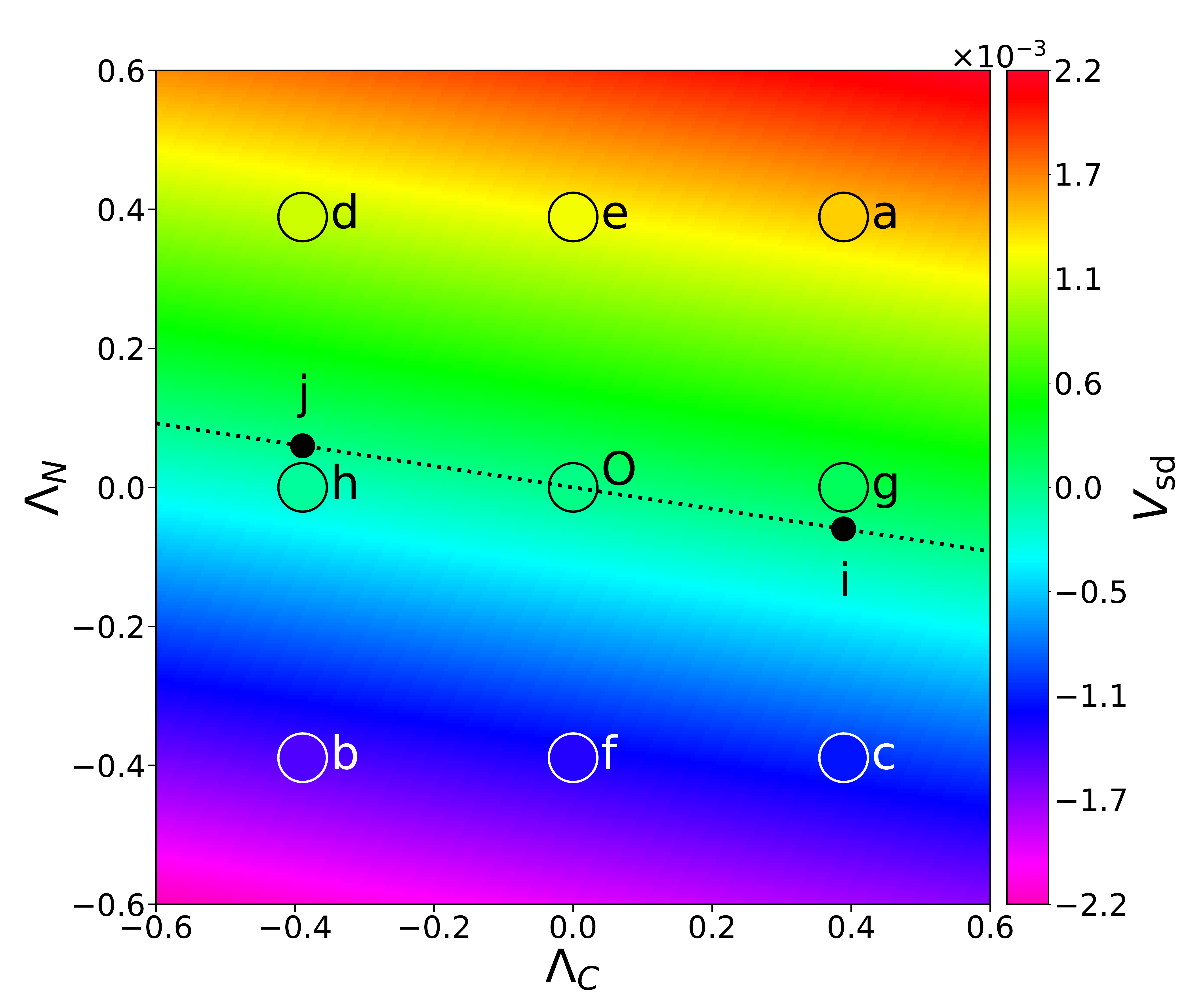}}
\caption{\label{fig:Lambda_map} The projection of the velocity along ${\bf u}$, $V_{\rm sd}$, of a single Janus motor operating in a reactive medium for various values of $\Lambda_C$ and $\Lambda_N$, where the catalytic reaction probabilities are $p_{\pm} =1$ and the bulk reaction rate constants are $k_2 = 5\times 10^{-5}$ and $k_{-2} = 5 \times 10^{-4}$. See Table~\ref{tab:Vu} for details.}
\end{figure}

When the fluid phase reaction $A \underset{k_{-2}}{\stackrel{k_2}{\rightleftharpoons}} B$ is present, the concentration fields are given by the solutions of the steady-state reaction-diffusion equations, $D_k \nabla^2 c_k +\nu_k w_2=0$, where $w_2= k_2 c_A- k_{-2} c_B$ is the bulk phase reaction rate. These equations must be solved subject to the radiation boundary conditions on the surface of the Janus colloid as described earlier, and concentrations $c_k=\bar{c}_k$ far from the colloid, which are determined by the fluid phase steady state conditions. Since $c_A+c_B=c_0$, the reaction-diffusion equation for $c_B$ takes the form, $(D\nabla^2 -k_2-k_{-2})c_B+k_2c_0=0$. The solution for $c_B$ has the same structure as Eq.~(\ref{eq:cb_solution}) but with $\bar{c}_B=k_2 c_0/(k_2+k_{-2})$ and the radial function $f_\ell(r)$ and coefficients given by the solutions of equations that account for the fluid phase reaction. Details are given in Appendix~\ref{app:conc}. The system is driven out of equilibrium by the condition $\kappa_+/\kappa_-=1 \neq k_2/k_{-2}=0.1$, thus breaking the detailed balance condition $\kappa_+/\kappa_-= k_2/k_{-2}$ that was considered in Ref.~[\onlinecite{HSGK18}] so that $(\kappa_+ \bar{c}_A - \kappa_-\bar{c}_B) > 0$.

\begin{figure}[htbp]
\centering
\resizebox{0.8\columnwidth}{!}{%
\includegraphics{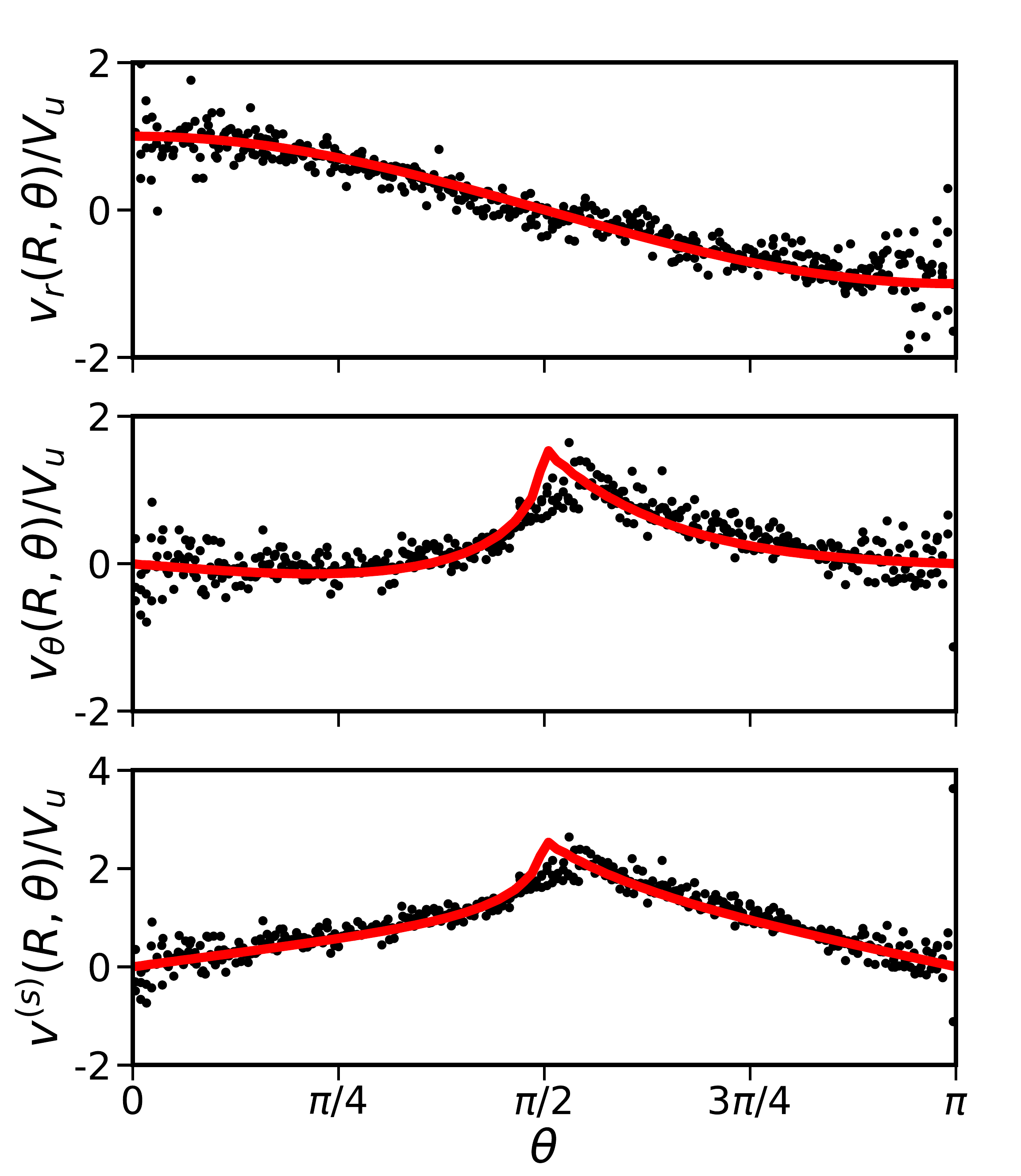}}
\caption{\label{fig:surface_velocities} Fluid velocity fields on the motor surface at point (a) of Fig.~\ref{fig:Lambda_map}.}
\end{figure}

The radial and tangential components of the velocity field at a position $R$ on the surface of the motor as well as the slip velocity ${\bf v}_{\rm slip}(R,\theta) = \hat{\bm{\theta}}[v_\theta(R,\theta)-\hat{\bm{\theta}}\cdot {\bf V}_{\rm sd}]$ are plotted in Fig.~\ref{fig:surface_velocities} and compared with the simulation results. The agreement is very good indicating that the continuum model with partial slip is able to accurately capture the forms of these fields obtained from direct simulations.

\begin{figure}[htbp]
\centering
\resizebox{0.7\columnwidth}{!}{%
\includegraphics{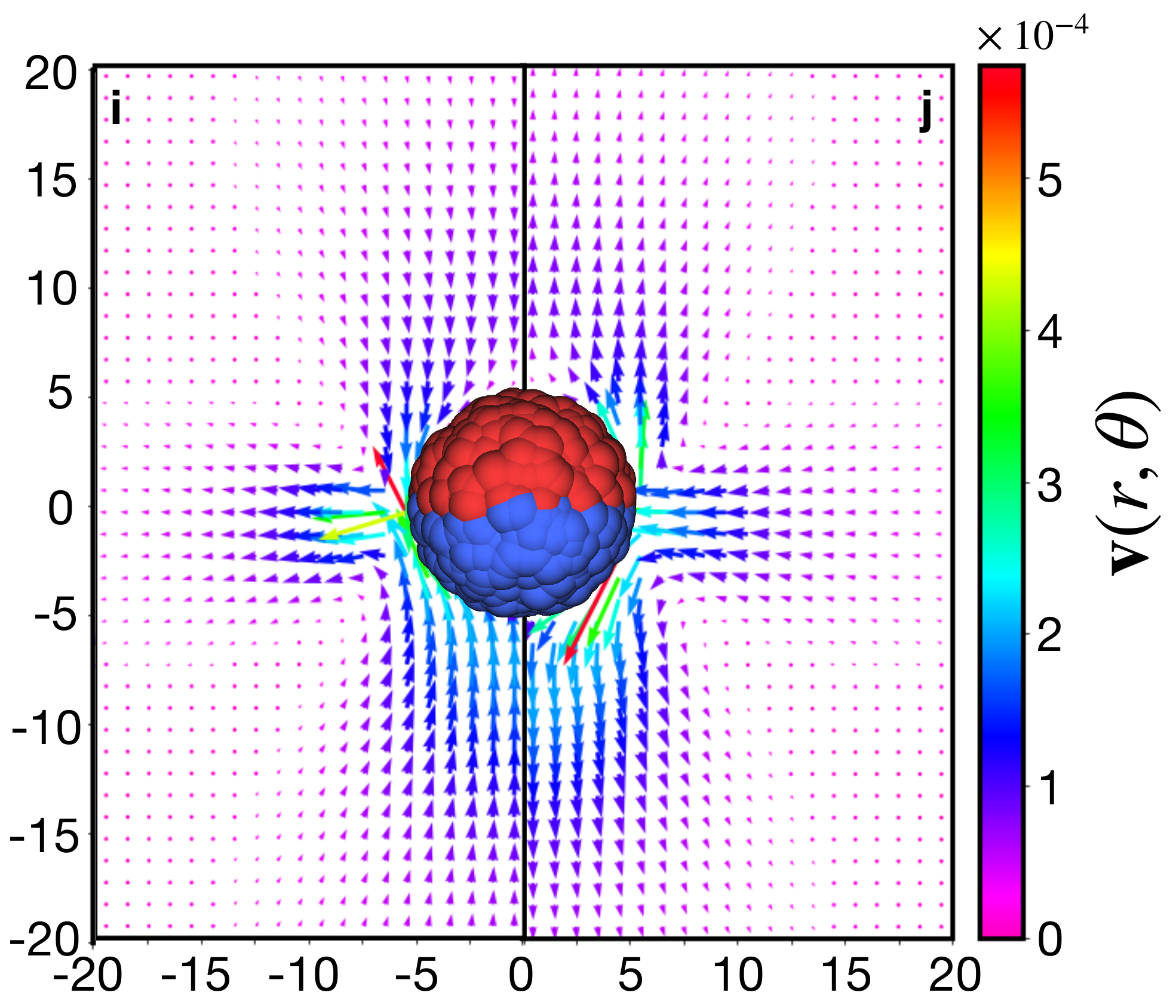}}
\caption{\label{fig:pump_theory}Fluid velocity fields in the vicinity of a Janus motor, where the corresponding $\Lambda$ factors are chosen to be points $i$ and $j$ on the dashed line shown in Fig.~\ref{fig:Lambda_map}.}
\end{figure}

The projection of the velocity along ${\bf u}$, $V_{\rm sd}$, is shown (color coded) in Fig.~\ref{fig:Lambda_map} as a function of the $\Lambda_C$ and $\Lambda_N$ parameters. In this figure the value of $V_{\rm sd}$ obtained from simulations (color inside small circles) is compared to the theoretical predictions computed using Eq.~(\ref{eq:Vu-cont}) (color outside circles) at selected points labeled (a)-(g). A quantitative comparison at these points in the $\Lambda_C$-$\Lambda_N$ plane is given in Table~\ref{tab:Vu}. Overall the agreement is generally within statistical errors.
\begin{table}
\caption{\label{tab:Vu} The motor propulsion velocities obtained from simulations ($S$) and theory ($T$) for various interaction strengths, $\epsilon_{H\alpha}$. The bulk reaction rates are $k_2 = 5\times 10^{-5}$ and $k_{-2} = 5\times 10^{-4}$. }
\begin{tabular}{cllcccccccccc}
\hline\hline
&	 &($\epsilon^C_{A},\epsilon^C_{B},\epsilon^N_{A},\epsilon^N_{B}$)& ($\Lambda_C,\Lambda_N$)&  $V_{\rm sd}^S\times 10^{3}$& $V_{\rm sd}^T\times 10^{3}$&\\
&(a)& (1.0,0.5,1.0,0.5)& (0.38,0.38)&  $1.41\pm 0.05$& $1.4$&\\
&(b)& (0.5,1.0,0.5,1.0)& (-0.38,-0.38)&  $-1.43\pm 0.06$& $-1.4$&\\
\\
&(c)& (1.0,0.5,0.5,1.0)& (0.38,-0.38)&  $-1.12\pm 0.04$& $-1.0$&\\
&(d)& (0.5,1.0,1.0,0.5)& (-0.38,0.38)&  $1.11\pm 0.06$& $1.0$&\\
\\
&(e)& (1.0,1.0,1.0,0.5)& (0.00,0.38)&  $1.22 \pm 0.06$& $1.2$&\\
&(f)& (1.0,1.0,0.5,1.0)& (0.00,-0.38)&  $-1.30 \pm 0.06$& $-1.2$&\\
\\
&(g)& (1.0,0.5,1.0,1.0)& (0.38,0.00)&  $0.16 \pm 0.05$& $0.18$&\\
&(h)& (0.5,1.0,1.0,1.0)& (-0.38,0.00)&  $-0.05 \pm 0.06$& $-0.18$&\\
\\
&(O)& (1.0,1.0,0.5,0.5)& (0.00,0.00)&  $0.01 \pm 0.05$& $0.0$&\\
\hline
\end{tabular}
\end{table}

The boundary that separates forward moving ($V_{\rm sd} > 0$) from backward moving ($V_{\rm sd} < 0$) motors is indicated by the dashed line where $V_{\rm sd} = 0$. For systems not in equilibrium, $\kappa_+ \bar{c}_A \ne \kappa_- \bar{c}_B$, a vanishing self-diffusiophoretic velocity implies $\Lambda_N = -\frac{\alpha_C}{\alpha_N} \Lambda_C$, where $\alpha_C$ and $\alpha_N$ are given in Eq.~(\ref{eq:alpha_h}). For the system parameters considered here $\alpha_C/\alpha_N \approx + 0.15$ . With the exception of the origin, along this line directed motion ceases although the diffusiophoretic mechanism is still in operation. Fluid flows with far-field dipolar character are generated as can be seen from the expressions for the fluid velocity field in Eqs.~(\ref{eq:vr}) and (\ref{eq:vt}) with $V_{\rm sd}=0$ and their plots in Fig.~\ref{fig:pump_theory}. Thus, along this line the diffusiophoretic forces arising from the catalytic and noncatalytic hemispheres have equal and opposite signs so that $V_{\rm sd}=0$ but fluid flows are still generated in the surrounding fluid.

\section{Collective behavior of many motors}\label{sec:collective}

Experimental and theoretical investigations of the collective behavior of active agents are currently being pursued on many different systems .~\cite{A13,EWG15,M16,ZS16,R17} The systems under study range from living macroscopic organisms such as fishes and birds to microscopic living entities such as bacteria, and natural or synthetic molecular machines and self-propelled motors. Theoretical models and associated simulations that focus on essential characteristics that are responsible for collective behavior in some situations have been proposed.~\cite{V12,CT13,BLS13,M16} Depending on the circumstances, the dynamical models may or may not include hydrodynamic coupling among the active agents. Typically, in all of these studies the goals are to understand the origins of nonequilibrium inhomogeneous collective states, how their appearance depends on system parameters, and the forms that they take. Investigations of collective behavior in active matter systems are important both from a fundamental perspective since they are examples of self-organization in active nonequilibrium systems which present challenges for theory, and from a practical perspective since most applications will involve the use of ensembles of motors rather than single motors.

Synthetic chemically-powered motors present additional features that affect their collective behavior. Since they move by phoretic mechanisms arising from self-generated concentration gradients, the inhomogeneous concentration fields in many-motor systems produce concentration gradient fields in the environment that can effect the motions of the motors, in addition to their self-propulsion. Thus, in such systems we must account for coupling of motors to one another through chemical gradients. In addition, phoretic propulsion is accompanied by fluid flow fields and hence hydrodynamic coupling will also play a role in the dynamics.  At high motor volume fractions direct motor-motor interactions also become important. There is a growing literature on theoretical and experimental studies of these systems.~\cite{IMS09,TCPYB12,BBKLBS13,PSSPC13,SGR14,WDAS15,GTLYBC15,PS15,LCM16,BM16}  Particle-based simulations of ensembles of hundreds Janus motors with variable catalytic cap sizes using a hard model and irreversible kinetics have been studied previously.~\cite{HSK17} Simulations with irreversible chemical reactions of the collective dynamics of small ensembles~\cite{TK12} and very large ensembles~\cite{CK17} with thousands of sphere-dimer motors~\cite{RK07,VTZKGKO10} have also been carried out.

In this section we present results of particle-based simulations of small ensembles of the Janus motors described in the previous section. In contrast to earlier investigations, we can study the effects of different intermolecular potentials for the catalytic and noncatalytic hemispheres on the collective dynamics  in a system whose dynamics satisfies detailed balance  and is maintained in a nonequilibrium steady state that is controlled by out-of-equilibrium fluid phase reactions.

\subsection{Simulation of collective dynamics}
We consider systems of $N_M = 20$ Janus motors in a cubic box of linear size $L$ with periodic boundary conditions. We primarily study dilute motor suspensions with volume fraction $\phi=V_{M}n_{M}/L^3= 0.05$, where $V_{M}$ is the volume of a motor.

\begin{figure}[htbp]
\centering
\resizebox{1.0\columnwidth}{!}{%
\includegraphics{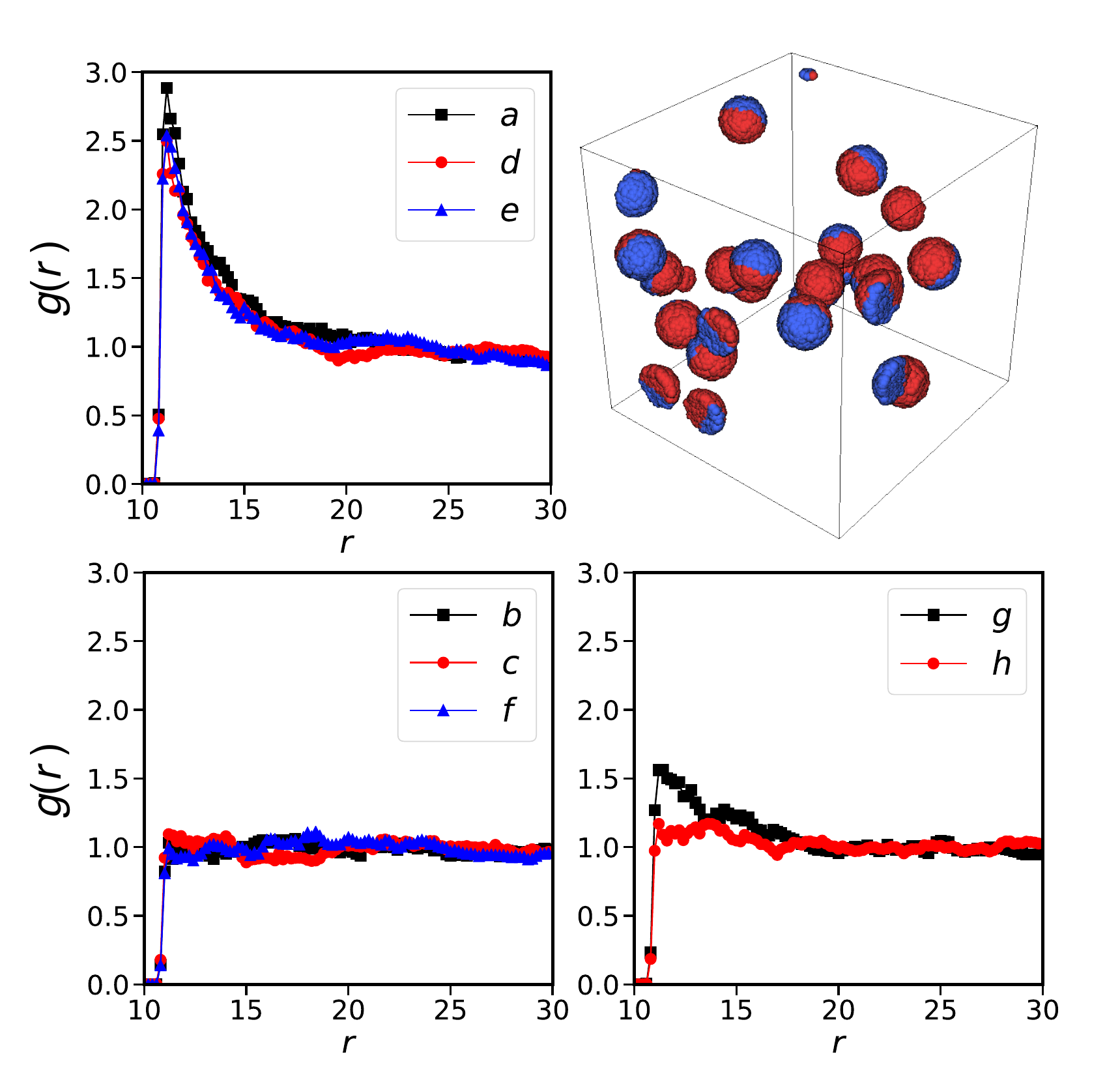}}
\caption{\label{fig:g_r}Radial distribution functions for the Janus motors with $\Lambda$ parameters listed in Table~\ref{tab:Vu}. In the simulation, the other parameters take the following values: $\kappa_+=\kappa_-=0.6$, $k_2=5\times 10^{-5}$, $k_{-2}=5\times 10^{-4}$, $n_A=9.1$, $n_B=0.9$, $a_0=0.0065$, $\Gamma=4\pi R^2 a_0=2.04$, $\alpha_C=-0.000993$, $\alpha_N=-0.006463$. The upper right panel shows an example of a dynamic cluster configuration with parameters for case (a).}
\end{figure}

The forms that the collective behavior take are first studied as function of the $\Lambda_h$ parameters listed in Table~\ref{tab:Vu}. To characterize the collective behavior, we consider the Janus motor radial distribution functions shown in Fig.~\ref{fig:g_r}. In this figure, the motors with $\Lambda_N > -\frac{\alpha_C}{\alpha_N} \Lambda_C$ show transient clustering (upper panels), whereas no clustering is seen in those cases with $\Lambda_N < -\frac{\alpha_C}{\alpha_N} \Lambda_C$ (lower-left panel). The lower-right panel compares the radial distribution functions $g(r)$ functions for the motors with propulsion arising only from diffusiophoretic effects from the $C$ hemispheres [cases (h) and (g) with $\Lambda_N=0$], and forward-moving motors [case (g), $\Lambda_C>0$] that show weak clustering. These results indicate that the $N$ hemisphere plays a dominant role in determining motor collective behavior. The average number of neighboring motors within first solvation shell ($r < r_s = 14$) from a given motor can be found from
\begin{equation}
\label{eq:ave_n}
\bar{n}(r_s) = 4 \pi \bar{\rho}_M \int_0^{r_s} g(r) r^2 dr,
\end{equation}
where $\bar{\rho}_M = N_M/L^3$ is the motor average density. We find that $\bar{n} \approx 2$ for cases (a, d, e) and $\bar{n} \approx 1$ for cases (b, c, f), indicating that forward-moving motors form clusters of average size $3$ and on average only dimers are found for backward-moving motors. While these average cluster numbers are small we note that the clusters are highly dynamic, forming and fragmenting, so that some system configurations have large clusters while others exhibit more gas-like configurations. The clustering ability of forward-moving motors and its lack for backward-moving motors is consistent with results obtained earlier for hard Janus particles~\cite{HSK17} and sphere-dimers~\cite{CK17}.

To further understand the structure of motor clusters and the effects of motor orientation, we consider the two-dimensional motor distribution function,
\begin{equation}
\rho_M (r,\theta) = \frac{1}{N_{M} \bar{\rho}_M}\sum_{i=1}^{N_{M}}\sum_{j\neq i}^{N_{M}} \delta(r_{ji}-r)\delta(\theta_{ji}- \theta),
\end{equation}
where $r_{ji} = |\mathbf{r}_{ji}|$ with $\mathbf{r}_{ji} = \mathbf{r}_j -\mathbf{r}_i$ is the distance between two motors and $\cos \theta_{ji} = \hat{\mathbf{u}}_i \cdot \mathbf{r}_{ji} / r_{ji} $, and the motor orientational function,
\begin{equation}
\mathbf{u}_M (r,\theta) = \frac{1}{N_{M} n_M(r,\theta)}\sum_{i=1}^{N_{M}}\sum_{j\neq i}^{N_{M}} \mathbf{u}_j' \delta(r_{ji}-r)\delta(\theta_{ji}- \theta)
\end{equation}
where $n_M(r,\theta) = N_{M}^{-1}\sum_{i=1}^{N_{M}}\sum_{j\neq i}^{N_{M}} \delta(r_{ji}-r)\delta(\theta_{ji}- \theta)$ is the average number of motors at $(r,\theta)$, and the components of $\mathbf{u}_j'$ are defined as
\begin{equation}
u_{jx}' =  \hat{\mathbf{u}}_j \cdot \frac{\mathbf{r}_{ji} - \mathbf{r}_{ji}\cdot \hat{\mathbf{u}}_i}{ |\mathbf{r}_{ji} - \mathbf{r}_{ji}\cdot \hat{\mathbf{u}}_i|}, \quad
u_{jy}' = \hat{\mathbf{u}}_j \cdot \hat{\mathbf{u}}_i.
\end{equation}

\begin{figure}[htbp]
\centering
\resizebox{0.8\columnwidth}{!}{%
\includegraphics{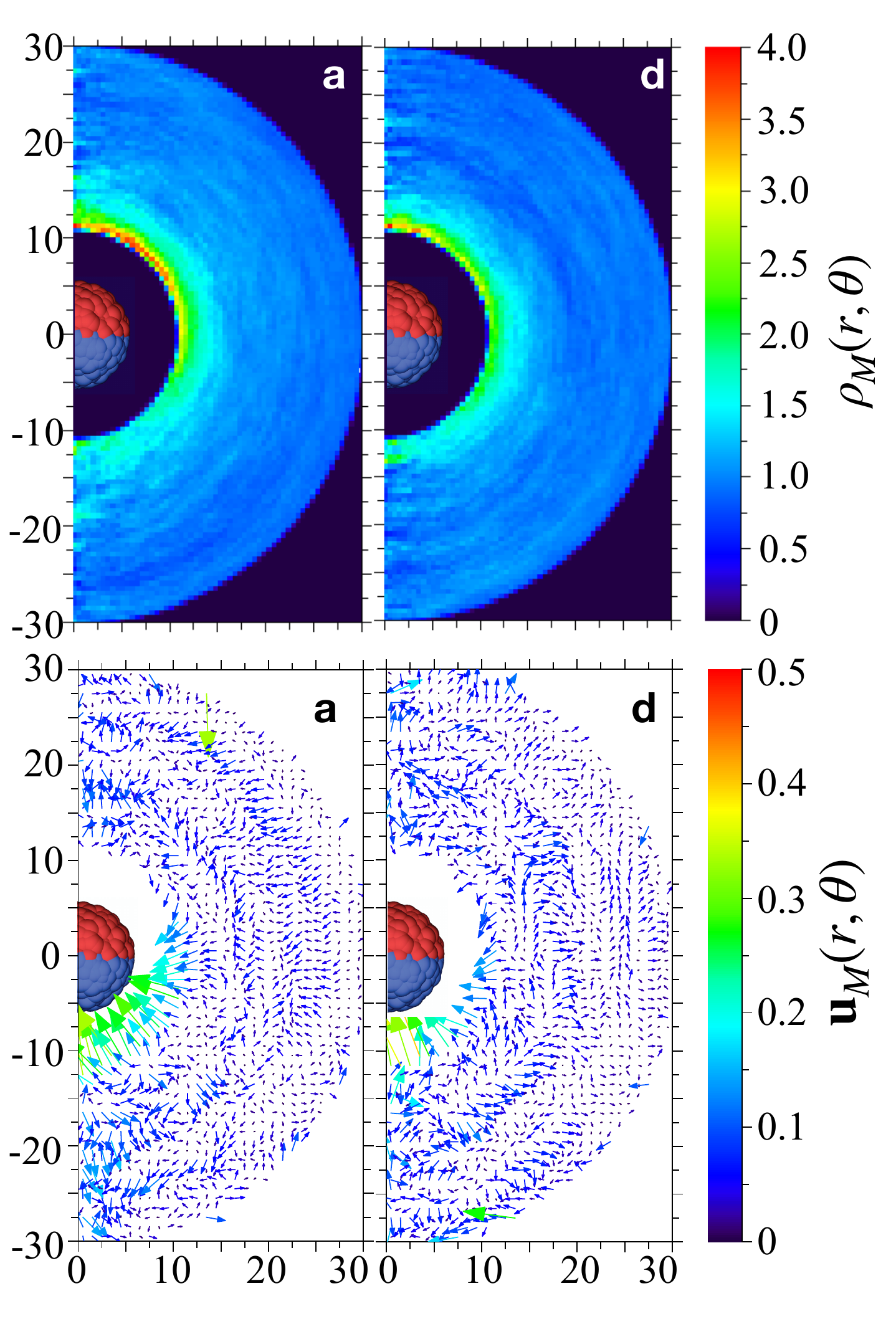}}
\caption{\label{fig:g_r_ori_compare_ab}Two dimensional motor density functions, $\rho_M (r,\theta)$, and motor orientation fields, $\mathbf{u}_M (r,\theta)$, for cases (a) and (d) listed in Table~\ref{tab:Vu}. For the orientation fields each arrow represents the average orientation of the motors at position ($r,\theta$) relative to the position and orientation of the reference motor.}
\end{figure}

\begin{figure}[htbp]
\centering
\resizebox{0.8\columnwidth}{!}{%
\includegraphics{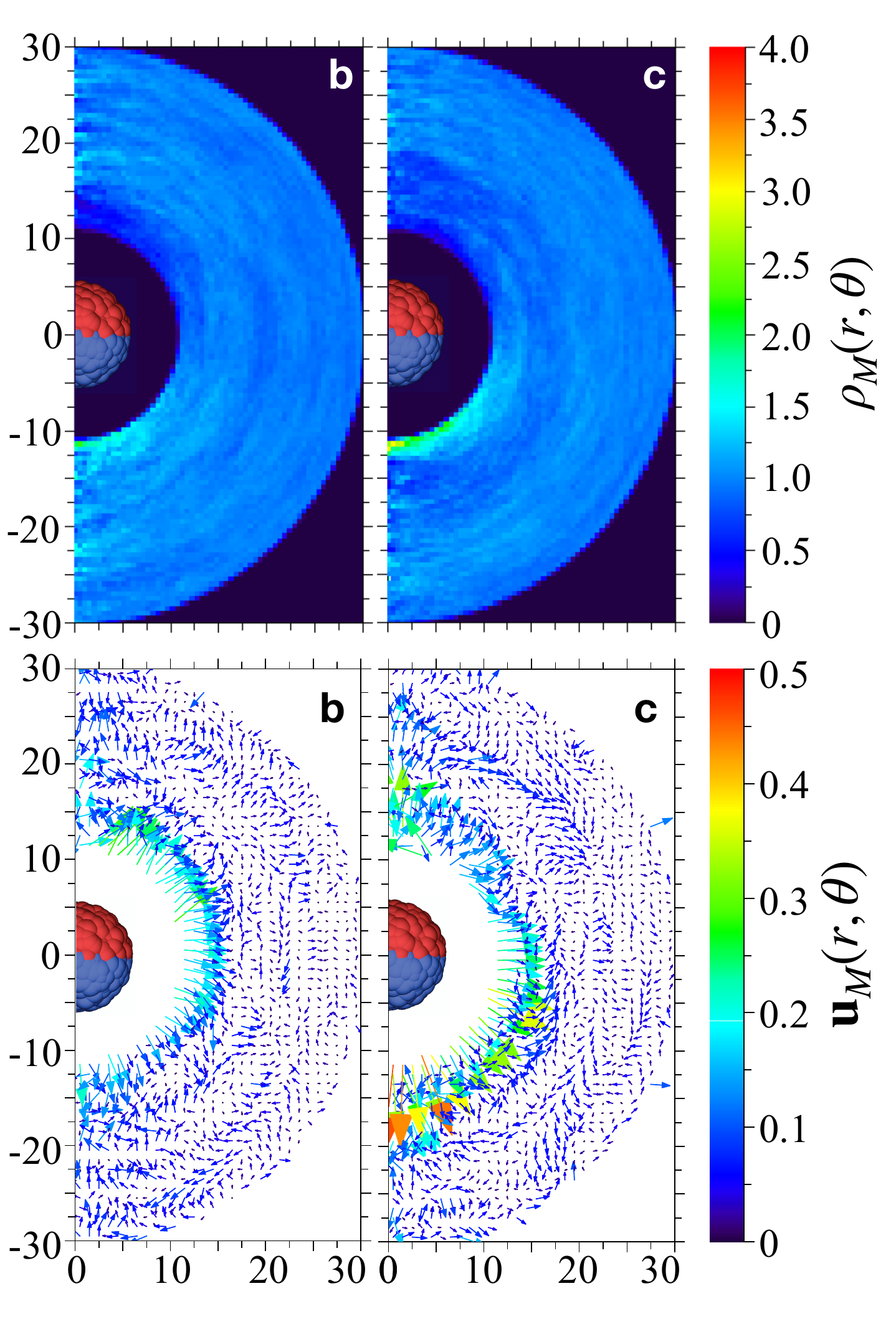}}
\caption{\label{fig:g_r_ori_compare_cd}Two dimensional motor density functions, $\rho_M (r,\theta)$,  and motor orientation fields, $\mathbf{u}_M (r,\theta)$, for cases (b) and (c) listed in Table~\ref{tab:Vu}.}
\end{figure}

\begin{figure}[htbp]
\centering
\resizebox{1.0\columnwidth}{!}{%
\includegraphics{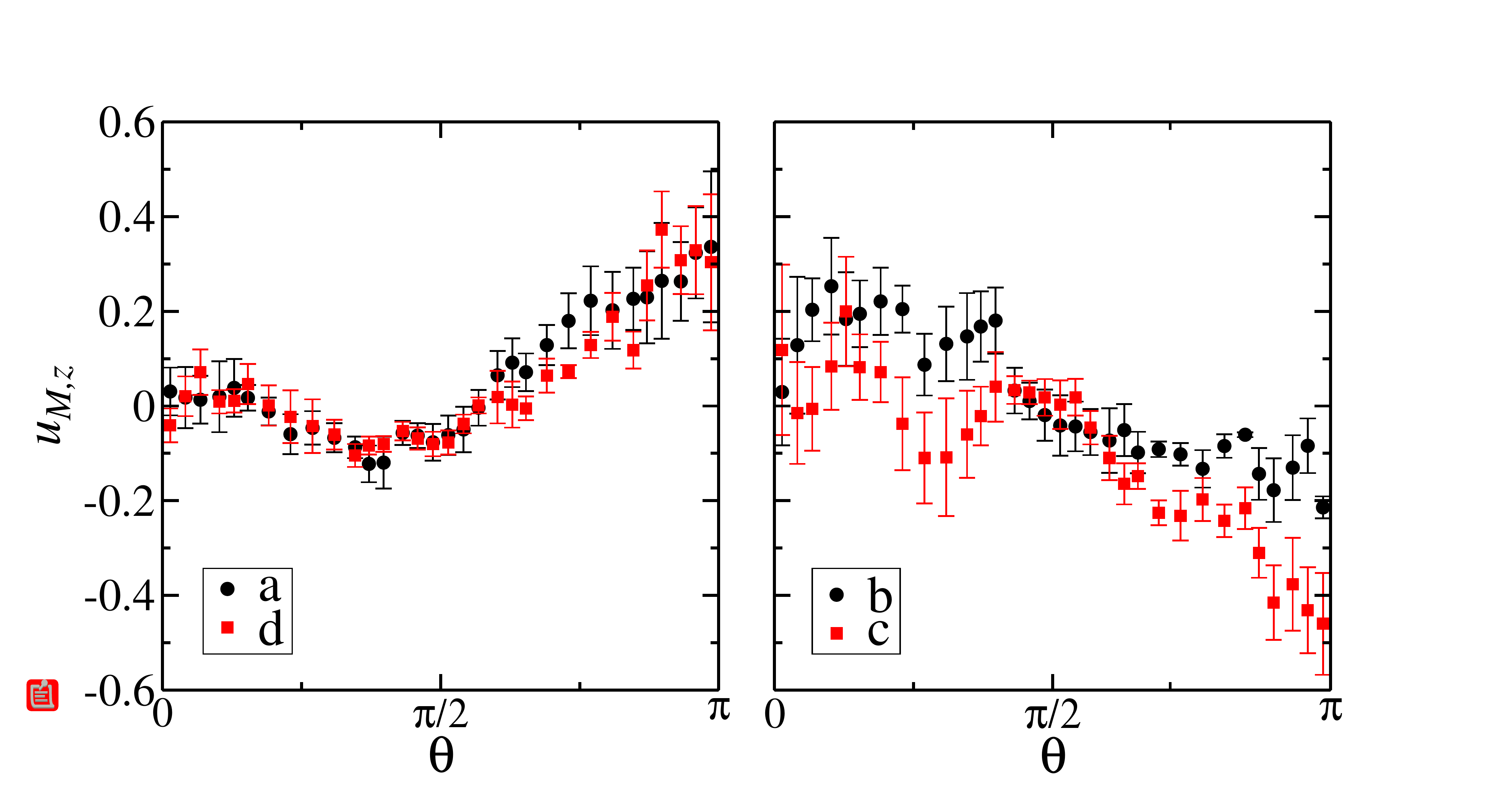}}
\caption{\label{fig:uM_first_shell_compare}Motor orientation fields in first solvation shell at $11 < r < 12$ (see Fig.~\ref{fig:g_r}) for forward-moving (left panel) and backward-moving (right panel) motors.}
\end{figure}

The results of calculations of $\rho_M (r,\theta)$ and $\mathbf{u}_M (r,\theta)$ are shown in Figs.~\ref{fig:g_r_ori_compare_ab} and \ref{fig:g_r_ori_compare_cd}, for forward-moving and backward-moving motors, respectively. For forward-moving motors, one sees that high/low motor densities and weak/strong orientational ordering in the vicinity of $C$/$N$ hemispheres are found. These observations suggest that the forward-moving motors tend to move toward each other and reside in regions with higher product concentrations near catalytic hemispheres. While only weak clustering is found for backward-moving motors, we notice that an enhanced clustering exists near the $N$ hemisphere for Janus motors with nonuniform $\Lambda$ parameters -- different $\Lambda$ values on the $C$ and $N$ hemispheres (Fig.~\ref{fig:g_r_ori_compare_cd}c, $\Lambda_C = -\Lambda_N > 0$). Such enhanced clustering can be explained by the existence of stronger orientational ordering. In contrast to the motors with uniform $\Lambda$ parameters (case b, $\Lambda_C = \Lambda_N < 0 $), where no net torque can be generated when interacting with the gradients of chemical species from other motors, motors with nonuniform $\Lambda$ parameters are able to adjust their orientation according to the direction of the gradient fields; i.e., the hemispheres $h$ with $\Lambda_h < 0$ prefer to move toward regions rich in $A$ particles, whereas those with $\Lambda_h > 0$ move away from $A$-rich regions. In case ($c$) for backward-moving motors, motors form pairs with $N$ hemispheres facing each other and, since reactions only occur on the $C$ hemisphere, the regions between $N$ hemispheres have higher $A$ concentration giving rise to a gradient of $A$ particles across the motors that stabilizes the pair configuration. Figure~\ref{fig:uM_first_shell_compare} shows the $z$ component of $\mathbf{u}_M (r,\theta)$ within the distance $ 11<r<12$ in the vicinity of the first solvation shell. Indeed, the average motor orientational ordering near the $N$ hemisphere ($\theta \approx \pi$) is stronger for case (c) than (b). Such an effect is not observed in simulations of forward-moving motors, since the gradient fields tend to average out when motors form clusters.

\subsection{Discussion}

\begin{figure}[htbp]
\centering
\resizebox{0.9\columnwidth}{!}{%
\includegraphics{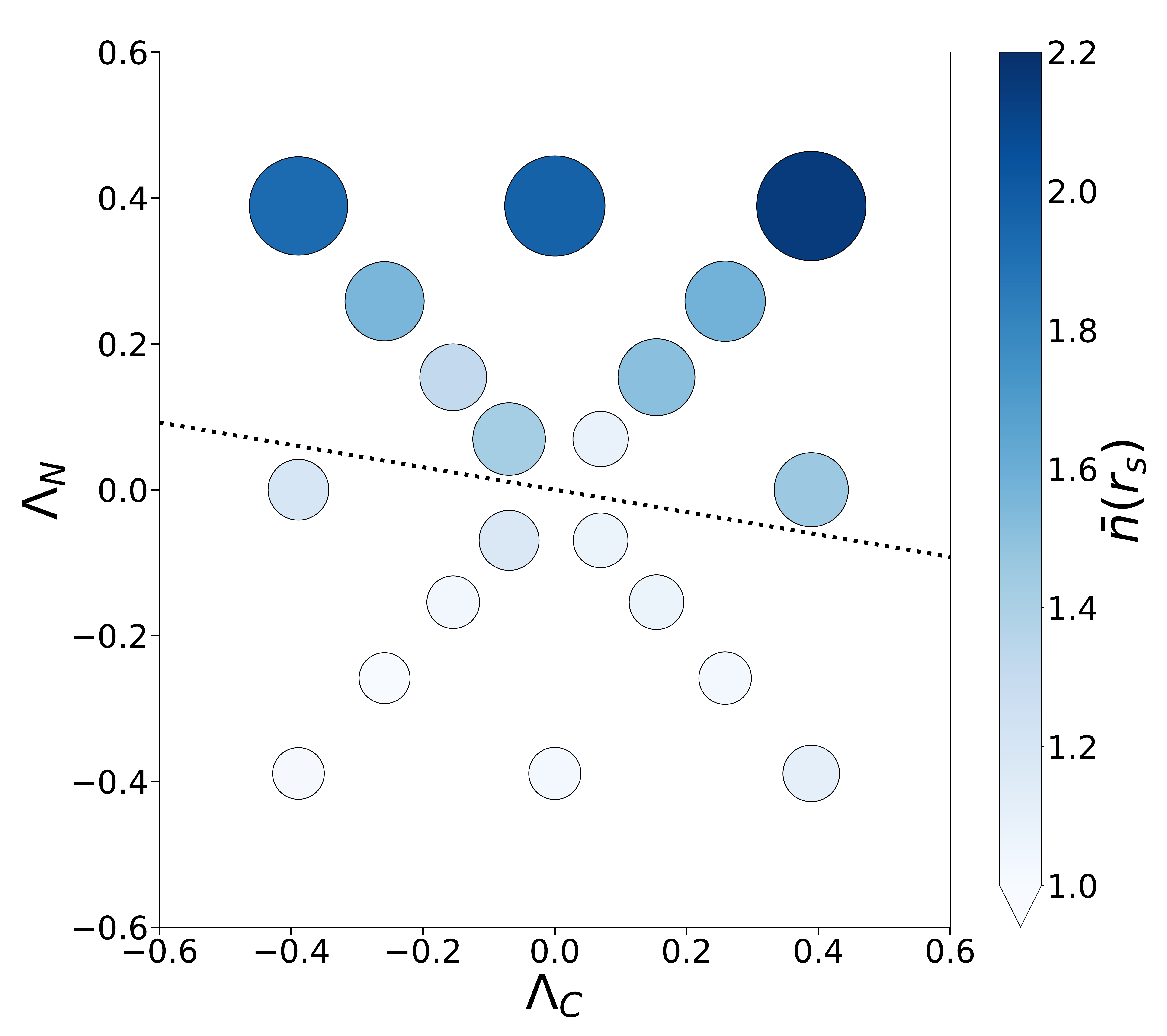}}
\caption{\label{fig:cluster_size} The diagram shows the degree of clustering for a selection of points in the $\Lambda_C -\Lambda_N$ plane. The scaled size and color gradation of each point indicate the number nearest-neighbor motors as determined from Eq.~(\ref{eq:ave_n}). Along the dotted $V_{\rm sd}=0$ where the motor velocity vanishes. System states that show dynamic clustering are seen for $V_{\rm sd}>0$.}
\end{figure}

As seen in Fig.~\ref{fig:g_r}, the radial distribution function $g(r)$ changes significantly depending on the sign of the self-diffusiophoretic velocity.  If $V_{\rm sd}>0$, the motors tend to move towards the fuel source and cluster together at short distances from each other so that $g(r)>1$ in the range of nearest-neighbors, $11<r<15$.  In contrast, if $V_{\rm sd}<0$, the motors tend to move away from the fuel source and they do not cluster, so that the radial distribution function remains close to unity $g(r)\simeq 1$ for $11<r<15$. These observations hold more generally as can be seen in Fig.~\ref{fig:cluster_size}, which presents more data in the $\Lambda_C -\Lambda_N$ plane. The size and color code of the points reflect the degree of clustering, and one can see that clusters tend to form in the region $V_{\rm sd}>0$. Note in particular, orientational effects are absent for points along the diagonal that have $\Lambda_C=\Lambda_N$, while these effects are present for points on the anti-diagonal, $\Lambda_C=-\Lambda_N$. The simulations show that motor-motor interactions play an important role in the observed clustering instability. Moreover, the threshold of the instability is essentially determined by the active self-diffusiophoretic velocity $V_{\rm sd}$, which thus plays a dominant role in the instability, as compared to passive diffusiophoresis.

The formation of small dynamic clusters in the present simulations is consistent with observations of cluster formation of Janus systems with hard interactions with $\Lambda_C=\Lambda_N$ where the effect of the size of the catalytic cap on cluster formation was studied.~\cite{HSK17} It was found that Janus motors with small catalytic caps showed the strongest tendency to form large stable clusters, while Janus motors with hemispherical caps showed a much weaker tendency to cluster, since the magnitude of the gradients on the Janus particle surface depends on the catalytic cap size.

\section{Summary and Conclusion} \label{sec:conc}

Several features of both the single-motor and collective results presented above are worth highlighting. All results were obtained using a particle-based dynamics that satisfies detailed balance, and the systems were driven out of equilibrium by controlling the chemical or mechanical affinities.

For single motors we showed that the reaction rate on a Janus motor, displaced from chemical equilibrium, has a quadratic dependence on a strong external force applied to the motor. This behavior lies outside the domain of the linear theory and our results should stimulate generalizations of the theory to explain this dependence. Furthermore, since the effect on the reaction rate is stronger than that in the linear regime, experimental investigations of external force effects on the reaction rate should be easier to observe. We also demonstrated that the finite-time fluctuation formula, which was derived for stationary catalytic particles, can also be applied to the Janus motors discussed in this paper. This result should prompt the development of extensions of the theory to treat chemically-powered motors. Lastly, by considering Janus motors with different interactions on the catalytic and noncatalytic hemispheres, regimes where the self-diffusiophoretic contributions from the different hemispheres can act in the same or different directions could be explored.

The studies of collective behavior were also carried out under well-defined nonequilibrium steady state conditions. While only small ensembles of motors were investigated, our results approximately determined the parameter domains under which the homogeneous motor state becomes unstable and leads to the formation of small dynamic clusters. The spatial and orientational features of these clusters were described in some detail. By considering systems with different $\Lambda_C$ and $\Lambda_N$ values the effects of self-diffusiophoresis and passive diffusiophoresis due to the presence of other motors could be studied. These results should also stimulate the development of theories that account for all of the features present in our coarse-grained microscopic simulations.

\appendix

\section{Simulation method and parameters}\label{app:sim}

The Janus motor is made from $N_m = 2681$ motor beads, each with mass $m$ and radius $\sigma$, randomly distributed within a sphere of radius $R_J = 4\:\sigma$ and the effective radius of the motor is $R = R_J+\sigma = 5\sigma$. The motor or motors and the $N = N_A + N_B + N_S$ particles of type $A$, $B$ and $S$ are in a cubic simulation box with linear size $L = 60\:\sigma$. The average densities of reactive and inert particles are $c_0 = (N_A+N_B) / L^3 = N_S / L^3 \approx 10/\sigma^3$. Multiparticle collision dynamics is implemented as described elsewhere~\cite{K08,GIKW09} with multiparticle collision time $\tau=0.1\:t_0$ with $t_0 = \sqrt{m \sigma^2/k_{\rm B}T}$; the molecular dynamics time step is $\delta t = 0.001\:t_0$. Fluid phase reactions are described by reactive multiparticle collision dynamics.\cite{RFK08} The motor-motor and motor-fluid interactions are determined by the repulsive Lennard-Jones potentials, $V_{HH}$ and $V_{\alpha H}$ (with $H=C,N$ and $\alpha=A,B,S$), respectively, with general form $V(r) =  4\epsilon [(\sigma'/r)^{12} - (\sigma'/r)^6 + 1/4]H(r_c-r)$, where $\sigma'$ is the interaction radius and $H(r_c-r)$ is a Heaviside function with $r_c = 2^{1/6}\sigma'$. The interaction strengths and radii are $\epsilon_{HH}=1.0$ and $\sigma_{HH} = 3\sigma$ for motor-motor interactions and $\epsilon_{SH}=0.5$ and $\sigma_{\alpha H} = \sigma$ for motor-solvent interactions and others as listed in Table~\ref{tab:Vu}.

Using the multiparticle collision expressions~\cite{K08,GIKW09} for the (common) solute diffusion constants $D$ and fluid viscosity $\eta$ one gets $D = 0.06$ and $\eta = 16.58$. The kinematic viscosity is $\nu = \eta/(2c_0) = 0.829$ and the Schmidt number is ${\rm Sc} = \nu/D =14$. The Janus colloid translational and rotational diffusion coefficients are $D_{\rm t} = 9\times 10^{-4}$ and $D_{\rm r} = 1.37\times 10^{-4}$, respectively. Partial slip conditions apply for the Janus particle and the slip length determined from these transport coefficients is $b \approx 10.5$. Here, we suppose that the slip length is the same on the catalytic and noncatalytic hemispheres: $b=b_C=b_N$. Further details are given in Ref.~[\onlinecite{HSGK18}].\\

\section{Concentration field}\label{app:conc}

The steady-state concentration field is obtained from the solution of the reaction-diffusion equation,
\begin{equation}
(\nabla^2 - \nu^2) c_B(r,\theta) + \frac{k_{2} c_0}{D} = 0,
\label{eq:rd_eq}
\end{equation}
with the notation $\nu^2\equiv (k_2+k_{-2})/D$, the boundary condition at large distance, $\lim_{r\to\infty} c_B(r,\theta) = c_0k_{2}/(k_2+k_{-2}) = \bar{c}_B$, and the radiation boundary condition on the surface of the Janus motor,
\begin{equation}\label{react-bc}
D \partial_r c_B(r,\theta) |_{r=R} = [\kappa_-c_B(R,\theta) - \kappa_+c_A(R,\theta)] H_C.
\end{equation}
Expressing the solution as an expansion in Legendre polynomials, $P_{\ell}(\mu)$, as  given in Eq.~(\ref{eq:cb_solution}),
we find that the radial function is given by
\begin{equation}
f_{\ell}(r) = \frac{\sqrt{\nu R}}{K_{\ell + \frac{1}{2}}(\nu R) } \frac{K_{\ell + \frac{1}{2}}(\nu r) }{\sqrt{\nu r}},
\label{eq:f_ell}
\end{equation}
where $K_{\ell}$ is a modified Bessel function. The coefficients $a_{\ell}$ can be found from the solution of a set of linear equations, $a_{\ell} =\sum_m (\mathbf{M}^{-1})_{\ell m} E_m$, where
\begin{equation}\label{matrix-M}
M_{\ell m } = \frac{2 Q_{\ell}}{2\ell + 1} \delta_{\ell m} + (\kappa_+ + \kappa_-)\frac{R}{D} \int_0^1 d\mu P_{\ell}(\mu) P_m(\mu) ,
\end{equation}
with $\mu=\cos\theta$, $Q_{\ell} = \nu R K_{\ell + \frac{3}{2}}(\nu R) / K_{\ell+\frac{1}{2}}(\nu R) -\ell$, and $E_m =\int_0^1 d\mu P_m(\mu)$.

In the absence of fluid phase reactions, we have $\nu=0$, and hence
\begin{equation}
\lim_{\nu\to 0} f_{\ell}(r)=(R/r)^{\ell +1} \quad \mbox{and} \quad \lim_{\nu\to 0} Q_{\ell}= \ell +1 \, ,
\end{equation}
thus recovering the results of Ref.~[\onlinecite{GK18a}].

\section{Fluid velocity field}\label{app:v_field}
The general solution of the Stokes equation $\nabla p = \eta \nabla^2 \mathbf{v}$ for an incompressible fluid $\nabla \cdot {\bf v} = 0$ with vanishing velocity far from the Janus colloid, $\lim_{r \to\infty} \mathbf{v} = 0$, can be written as~\cite{RHSK16}
\begin{eqnarray}
v_r(r,\theta) &=& \sum_{\ell = 1}^{\infty} \bigg[\frac{\ell +1 }{ 2\eta (2\ell -1) }\bigg(\frac{R}{r}\bigg)^{\ell} \frac{\chi_{\ell}^{(1)}}{R^{\ell}} \nonumber \\
&&\quad - (\ell+1)\bigg( \frac{R}{r} \bigg)^{\ell+2}\frac{\chi_{\ell}^{(2)}}{R^{\ell + 2}}  \bigg] P_{\ell}(\cos \theta), \label{eq:gen_solu_vr}\\
v_{\theta}(r,\theta) &=& \sum_{\ell = 1}^{\infty} \bigg[\frac{2 - \ell }{2 \eta \ell (2\ell-1)}\bigg(\frac{R}{r}\bigg)^{\ell} \frac{\chi_{\ell}^{(1)}}{R^{\ell}} \nonumber \\
&&\quad + \bigg( \frac{R}{r} \bigg)^{\ell+2}\frac{\chi_{\ell}^{(2)}}{R^{\ell + 2}}  \bigg] P_{\ell}^1 (\cos \theta) \label{eq:gen_solu_vt}.
\end{eqnarray}
The coefficients $\chi_{\ell}^{(1,2)}$ can be determined by using the boundary conditions at $R$:
\begin{eqnarray}
v_r(R, \theta) &=& {\bf V}_{\rm sd} \cdot \hat{{\bf r}}, \label{eq:bc_vr} \\
v_\theta(R, \theta)&-&{\bf V}_{\rm sd}\cdot \hat{\bm{\theta}} =
 b\left(\partial_r v_\theta+\partial_{\theta} v_r/r-v_{\theta}/r\right)_{r=R} \nonumber\\
&& -\frac{k_{\rm B}T}{\eta R} \, \partial_{\theta}c_B(R,\theta)\sum_{h=C}^N \Lambda_h H_h ,  \label{eq:vs}
\end{eqnarray}
where the tangential gradient of the concentration field is given by
\begin{equation}
\frac{1}{R}\, \partial_{\theta}c_B(\bar{R},\theta) = (\kappa_+ \bar{c}_A - \kappa_- \bar{c}_B)\frac{1}{D} \sum_{\ell = 1}^{\infty} a_{\ell} P_{\ell}^1(\cos \theta).
\label{eq:grad_cB}
\end{equation}
Since ${\bf V}_{\rm sd} \cdot \hat{{\bf r}}=V_{\rm sd} P_1(\cos \theta)$, comparing Eqs.~(\ref{eq:gen_solu_vr}) and (\ref{eq:bc_vr}) for $\ell =1 $ and $\ell \geq 2$ yields the equations
\begin{equation}
\frac{1}{\eta} \frac{\chi_{1}^{(1)}}{R} = V_{\rm sd} + 2\, \frac{\chi_{1}^{(2)}}{R^3}, \quad
\frac{1}{2 \eta (2 \ell-1) } \frac{\chi_{\ell}^{(1)}}{R^{\ell}} =\frac{\chi_{\ell}^{(2)}}{R^{\ell+2}} \label{eq:coef_relation},
\end{equation}
respectively. Substituting Eqs.~(\ref{eq:coef_relation}) into Eqs.~(\ref{eq:gen_solu_vr}) and (\ref{eq:gen_solu_vt}), noting that ${\bf u} \cdot \hat{\bm{\theta}}= P_1^1(\cos \theta)$, and replacing coefficient by $\chi_{\ell}^{(2)}/ R^{\ell +2} = \chi_{\ell} $ yields
\begin{eqnarray}
v_r(r, \theta) &=& V_{\rm sd}\bigg( \frac{R}{r}\bigg) P_1(\cos \theta) \\
&&+ \sum_{\ell=1}^{\infty} (\ell + 1)\bigg[ \bigg( \frac{R}{r}\bigg)^{\ell}  - \bigg(\frac{R}{r}\bigg)^{\ell+2} \bigg] \chi_{\ell} P_{\ell} (\cos \theta), \label{eq:v_r_tmp} \nonumber\\
v_{\theta}(r, \theta) &=& V_{\rm sd}\bigg( \frac{R}{2 r}\bigg) P_1^1(\cos \theta) \\
&&+ \sum_{\ell=1}^{\infty} \bigg[\bigg( \frac{2-\ell}{\ell}\bigg) \bigg( \frac{R}{r}\bigg)^{\ell} + \bigg(\frac{R}{r}\bigg)^{\ell+2} \bigg] \chi_{\ell} P_{\ell}^1 (\cos \theta) \label{eq:v_t_tmp} \nonumber.
\end{eqnarray}

Using the boundary condition in the tangential direction (Eq.~(\ref{eq:vs})), the coefficient $\chi_1$ is found to be
\begin{equation}
\chi_{1} = \frac{3}{4}\bigg[ \frac{V_{\rm sd}}{3(1+3b/R)} - \frac{1+2b/R}{1+3b/R}\sum_{m=1}^{\infty}B_{1 m} a_{m} \bigg],
\end{equation}
and $\chi_{\ell}$ with $\ell \geq 2$ as given in the main text with $B_{\ell m}$ in Eq~(\ref{eq:B_lm}). Evaluating the sum $\sum_{m=1}^{\infty}B_{1 m} a_{m} = V_{\rm sd}$ through the comparison with Eq.~(\ref{eq:Vu-cont}) so that $\chi_1 = -V_{\rm sd}/2$, Eqs.~(\ref{eq:v_r_tmp}) and (\ref{eq:v_t_tmp}) can be rewritten in the form of the equations~(\ref{eq:vr}) and (\ref{eq:vt}) for $v_r(r,\theta)$ and $v_{\theta}(r,\theta)$ given in the main text.

\section*{Acknowledgments}
Research was supported in part by the Natural Sciences and Engineering Research Council of Canada and Compute Canada. Financial support from the International Solvay Institutes for Physics and Chemistry, the Universit\'e libre de Bruxelles (ULB), the Fonds de la Recherche Scientifique~-~FNRS under the Grant PDR~T.0094.16 for the project ``SYMSTATPHYS" is also acknowledged.

%

\end{document}